\documentclass[article]{IEEEtran}
\IEEEoverridecommandlockouts
\usepackage{mdwmath}
\usepackage{mdwtab}

\usepackage{amsfonts}
\usepackage{amsmath}
\usepackage{amsthm}
\usepackage{amssymb}
\usepackage{graphicx}
\usepackage{subcaption}
\usepackage{breqn}
\usepackage{setspace}
\usepackage[T1]{fontenc}
\usepackage{supertabular}
\usepackage{longtable}
\usepackage{bbm}
\usepackage{url}
\usepackage{enumerate}
\usepackage{caption}
\usepackage{fancyhdr}
\usepackage{breqn}
\usepackage{fixltx2e}
\usepackage{capt-of}
\usepackage{mdframed}
\newcommand{\mathsym}[1]{{}}
\newcommand{\unicode}[1]{{}}

\begin{document}

\title{On the statistical properties and tail risk of violent conflicts}

\author{ 
\IEEEauthorblockN{Pasquale Cirillo\IEEEauthorrefmark{1},
Nassim Nicholas Taleb\IEEEauthorrefmark{2}\\}
\IEEEauthorblockA{\IEEEauthorrefmark{1}Applied Probability Group, Delft University of Technology,
\IEEEauthorrefmark{2}Tandon School of Engineering, New York University} 
}




\maketitle
\begin{abstract}

We examine statistical pictures of violent conflicts over the last 2000 years, finding techniques for dealing with incompleteness and unreliability of historical data. 

We introduce a novel approach to apply extreme value theory to fat-tailed variables that have a remote, but nonetheless finite upper bound, by defining a corresponding unbounded dual distribution (given that potential war casualties are bounded by the world population). 

We apply methods from extreme value theory on the dual distribution and derive its tail properties. The dual method allows us to calculate the real mean of war casualties, which proves to be considerably larger than the sample mean, meaning severe underestimation of the tail risks of conflicts from naive observation. We analyze the robustness of our results to errors in historical reports, taking into account the unreliability of accounts by historians and absence of critical data. 

We study inter-arrival times between tail events and find that no particular trend can be asserted. 

All the statistical pictures obtained are at variance with the prevailing claims about "long peace", namely that violence has been declining over time.
\end{abstract} 
\thispagestyle{fancy} 

\markboth{\textbf{Tail Risk Working Papers}}
\flushbottom

\section{Introduction}
Since the middle of last century, there has been a multidisciplinary interest in  wars and armed conflicts (quantified in terms of casualties), see for example \cite{Berlinski}, \cite{Goldstein}, \cite{Mueller}, \cite{Pinker}, \cite{Richardson2}, \cite{Richardson}, \cite{Wallensteen} and \cite{White1}.  Studies have also covered the statistics of terrorism, for instance \cite{Clauset}, \cite{Scharpf}, and the special issue of \textit{Risk Analysis} on terrorism \cite{RiskAnalysis}. From a statistical point of view, recent contributions have attempted to show that the distribution of war casualties (or terrorist attacks' victims) tends to have heavy tails, characterized by a power law decay \cite{Clauset} and \cite{Friedman}. Often, the analysis of armed conflicts falls within the broader study of violence \cite{Cederman}, \cite{Pinker}, with the aim to understand whether we as human are more or less violent and aggressive than in the past and what role institutions played in that respect.  Accordingly, the public intellectual arena has witnessed active debates, such as the one between Steven Pinker on one side, and John Gray on the other concerning the hypothesis that the long peace was a statistically established phenomenon or a mere statistical sampling error that is characteristic of heavy-tailed processes, \cite{Gray} and \cite{Norton} --the latter of which is corroborated by this paper. 

Using a more recent data set containing 565 armed conflicts with more than 3000 casualties over the period 1-2015 AD, we confirm that the distribution of war casualties exhibits a very heavy right-tail. The tail is so heavy that --- at  first glance --- war casualties could represent an infinite-mean phenomenon, as defined by \cite{Neslehova}. But should the distribution of war casualties have an infinite mean, the annihilation of the human species would be just a matter of time, and the sample properties we can compute from data have no meaning at all in terms of statistical inference. In reality, a simple argument allows us to rule out the infiniteness of the mean: no  event or series of events can kill more than the whole world population. The support of the distribution of war casualties is necessarily bounded, and the true mean cannot be infinite. 

Let $[L,H]$ be the support of the distribution of war casualties today. $L$ cannot be smaller than 0, and we can safely fix it at some level $L^\ast>>0$ to ignore those small events that are not readily definable as armed conflict \cite{Wallensteen}. As to $H$, its value cannot be larger than the world population, i.e. 7.2 billion people in 2015\footnote{Today's world population can be taken as the upper bound, as never before humanity reached similar numbers.} \cite{UN}.

If $Y$ is the random variable representing war casualties, its range of variation is very large and equal to $H-L$.  Studying the distribution of $Y$ can be difficult, given its bounded but extremely wide support. Finding the right parametric model among the family of possible ones is hard, while nonparametric approaches are more difficult to interpret from a risk analysis point of view. 

Our approach is to transform the data in order to apply the powerful tools of extreme value theory.  Since $H<\infty$ we suggest a log-transformation of the data.  This allows to use tools such as  \cite{deHaan}. 
 Theoretical results such as the Pickands, Balkema and de Haan's Theorem allow us to simplify the choice of the model to fit to our data.


\begin{figure*}[h!]
\includegraphics[width=\linewidth]{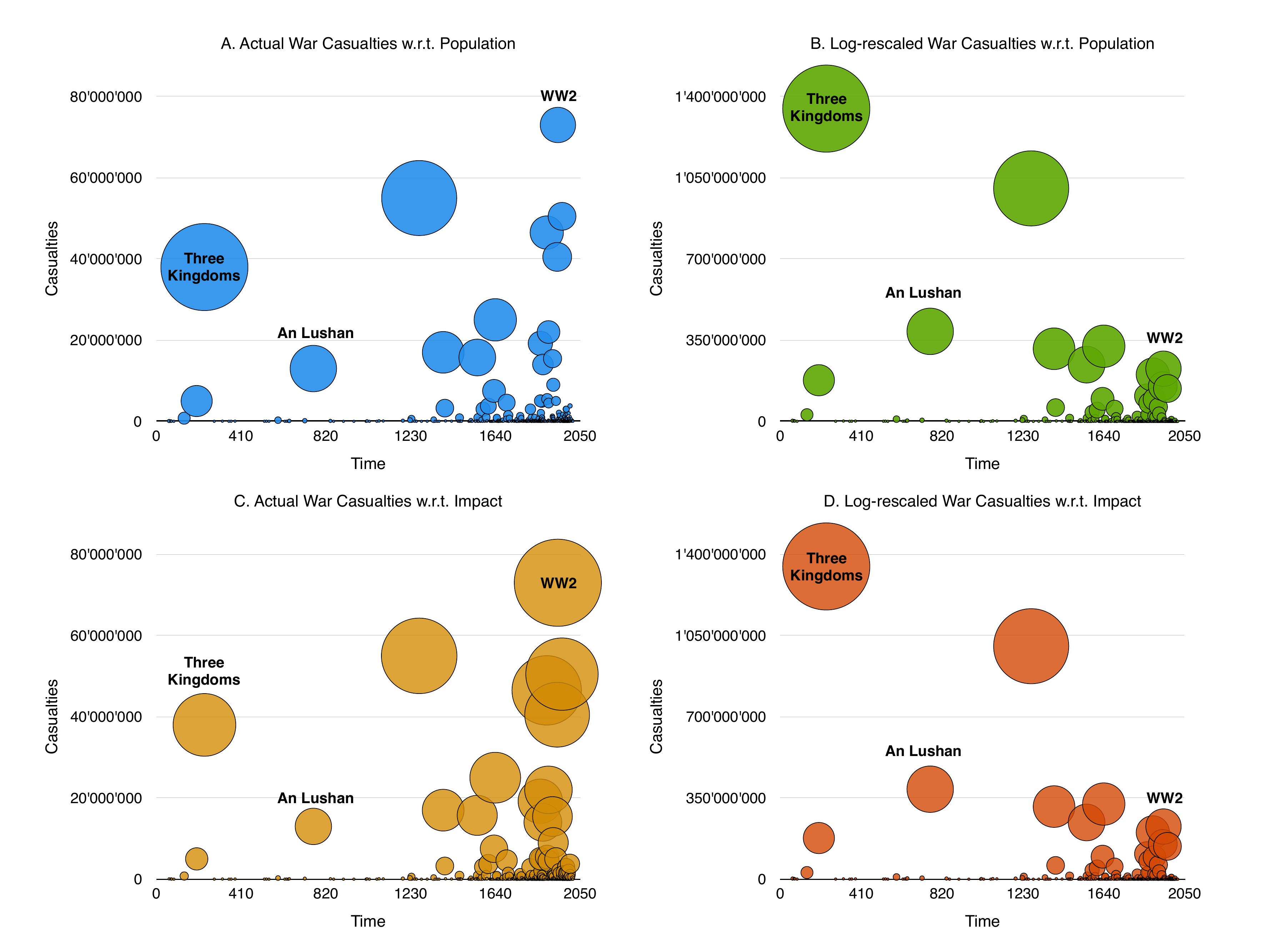}
\caption{War casualties over time, using raw (A,C) and dual (B,D) data. The size of each bubble represents the size of each event with respect to today's world population (A,B) and with respect to the total casualties (raw: C, rescaled: D) in the data set.}
\label{bubbles}
\end{figure*}

Let $L$ and $H$ be respectively the lower and the upper bound of a random variable Y, and define the function
\begin{equation}
\varphi(Y)= 
L-H \log \left(\frac{H-Y}{H-L}\right).
\label{transform}
\end{equation}
It is easy to verify that
\begin{enumerate}
\item $\varphi$  is "smooth": $\varphi \in C^\infty$,
\item $\varphi^{-1}(\infty)=H,$
\item $\varphi^{-1}(L)=\varphi(L)=L$.
 \end{enumerate} 
Then $Z=\varphi(Y)$ defines a new random variable with lower bound $L$ and an infinite upper bound. Notice that the transformation induced by $\varphi(\cdot)$ does not depend on any of the parameters of the distribution of $Y$. In what follows, we will call the distributions of $Y$ and $Z$, respectively the real and the dual distribution.

By studying the tail properties of the dual distribution (the one with an infinite upper bound), using extreme value theory, we will be able to obtain, by reverting to the real distribution, what we call the \textit{shadow} mean of war casualties. We will show that this mean is at least 1.5 times larger than the sample mean, but nevertheless finite.

We assume that many observations are missing from the dataset (from under-reported conflicts), and we base on analysis on the fact that war casualties are just imprecise estimates \cite{Spagat}, on which historians often have disputes, without anyone's  ability to verify the assessments using period sources. For instance, an event that took place in the eighth century, the An Lushan rebellion, is thought to have killed 13 million people, but there no precise or reliable methodology to allow us to trust that number --which could be conceivably one order of magnitude smaller.\footnote{For a long time, an assessment of the drop in population in China was made on the basis of tax census, which might be attributable to a drop in the aftermath of the rebellion in surveyors and tax collectors.\cite{BBC}}. Using resampling techniques, we show that our tail estimates are robust to changes in the quality and reliability of data. Our results and conclusions will replicate even we missed a third of the data.
 When focusing on the more reliable set covering the last 500 years of data, one cannot observe any specific trend in the number of conflicts, as large wars appear to follow a homogeneous Poisson process. This memorylessness in the data conflicts with the idea that war violence has declined over time, as proposed by \cite{Goldstein} or \cite{Pinker}.

The paper is organized as follows: Section \ref{data} describes our data set and analyses the most significant problems with the quality of observations; Section \ref{danalysis} is a descriptive analysis of data; it shows some basic result, which are already sufficient to refute the thesis \cite{Goldstein} that we are living in a more peaceful world on the basis of statistical observations; Section \ref{eanalysis} contains our investigations about the upper tail of the dual distribution of war casualties as well as discussions of tail risk; Section \ref{ghost} deals with the estimation of the shadow mean; in Section \ref{missingdata} we discuss the robustness of our results to imprecision and errors in the data; finally, Section \ref{trendarm} looks at the number of conflicts over time, showing no visible trend in the last 500 years.

\section{About the data} \label{data}
The data set contains 565 events over the period 1-2015 AD; an excerpt is shown in Table \ref{data}. Events are generally armed conflicts\footnote{We refer to the definition of \cite{Wallensteen}, according to which ``An armed conflict is a contested incompatibility which concerns government and/or territory where the use of armed force between two parties, of which at least one is the government of a state, results in at least 25 battle-related deaths". This definition is also compatible with the Geneva Conventions of 1949.}, such as interstate wars and civil wars, with a few exceptions represented by violence against citizens perpetrated by the bloodiest dictatorships, such as Stalin's and Mao Zedong's regimes. These were included in order to be consistent with previous works about war victims and violence, e.g. \cite{Pinker} and \cite{Richardson}. 

We had to deal with the problem of inconsistency and lack of uniformity in the attribution of casualties by historians. Some events such as the siege of Jerusalem include death from famine, while for other wars only direct military victims are counted.  It might be difficult to disentangle death from direct violence from those arising from such side effects as contagious diseases, hunger, rise in crime, etc. Nevertheless, as we show in our robustness analysis  these inconsistencies do not affect out results ---in contrast with analyses done on thin-tailed data (or data perceived to be so) where conclusions can be reversed on the basis of a few observations.

The different sources for the data are: \cite{Berlinski}, \cite{Hayek}, \cite{Necro}, \cite{Phillips}, \cite{Pinker}, \cite{Richardson}, \cite{Wallensteen}, \cite{White1} and \cite{White2}. For websites like Necrometrics \cite{Necro}, data have been double-checked against the cited references. The first observation in our data set is the Boudicca's Revolt of 60-61 AD, while the last one is the international armed conflict against the Islamic State of Iraq and the Levant, still open.

\begin{table*}[h!]
\small{
\caption{Excerpt of the data set of war casualties. The original data set contains 565 events in the period 1-2015 AD. For some events more than one estimate is available for casualties and we provide: the minimum (Min), the maximum (Max), and an intermediate one (Mid) according to historical sources. Casualties and world population estimates (Pop) in 10000.}
\begin{center}
\begin{tabular}{|l|c|c|c|c|c|c|}
\hline
\textbf{Event} &\textbf{Start}& \textbf{End} &\textbf{Min} & \textbf{Mid} & \textbf{Max} & \textbf{Pop} \\
\hline
Boudicca's Revolt& 60 & 61 & 70& 7.52 &8.04& 19506\\
Three Kingdoms & 220 & 280 & 3600& 3800 &4032 & 20231\\
An Lushan's Reb.& 755 & 763 & 800& 1300& 3600& 24049\\
Sicilian Vespers & 1282 & 1302 & - & 0.41& 0.80& 39240\\
WW1 & 1914 & 1918 & 1466& 1544& 1841& 177718\\
WW2 & 1939 & 1945 & 4823& 7300& 8500& 230735\\
\hline
\end{tabular}
\end{center}
\label{data}}
\end{table*}

We include all events with more than 3000 casualties (soldiers and civilians) in absolute terms, without any rescaling with respect to the coeval world population. More details about rescaling in Subsection \ref{rescaling}.  The choice of the 3000 victims threshold was motivated by three main observations:
\begin{itemize}
\item Conflicts with a high number of victims are more likely to be registered and studied by historians. While it is easier today to have reliable numbers about minor conflicts --- although this point has been challenged by \cite{Seybolt} and \cite{Spagat} --- it is highly improbable to ferret out all smaller events that took place in the fifth century AD. In particular, a historiographical bias prevents us from accounting for much of the conflicts that took place in the Americas and Australia, before their discovery by European conquerors.
\item A higher threshold gives us a higher confidence about the estimated number of casualties, thanks to the larger number of sources,  even if, for the bloodiest events of the far past, we must be careful about the possible exaggeration of numbers.
\item The object of our concern is tail risk. The extreme value techniques we use to study the right tail of the distribution of war casualties imposition of thresholds, actually even larger than 3000 casualties. 
\end{itemize}
To rescale conflicts and expressing casualties in terms of today's world population (more details in Subsection \ref{rescaling}), we relied on the population estimates of \cite{Klein} for the period 1-1599 AD, and of \cite{UN6B} and \cite{UN} for the period 1650-2015 AD. We used:
\begin{itemize}
\item Century estimates from 1 AD to 1599. 
\item Half-century estimates from 1600 to 1899.
\item Decennial estimates from 1900 to 1949.
\item Yearly estimates from 1950 until today.
\end{itemize}

\subsection{Data problems} \label{probl}
Accounts of war casualties are often anecdotal, spreading via citations, and based on vague estimates, without anyone's ability to verify the assessments using period sources. 
 For instance, the independence war of Algeria has various estimates, some from the French Army, others from the rebels, and nothing scientifically obtained \cite{Horne}. 

This can lead to several estimates for the same conflict, some more conservative and some less. Table \ref{data}, shows different estimates for most conflicts. In case of several estimates, we present the minimum, the average and the maximum one. Interestingly, as we show later, choosing one of the three as the ``true" estimate does not affect our results --thanks to the scaling properties of power laws.

 Conflicts, such as the Mongolian Invasions, which we refer to as ``named" conflicts, need to be treated with care from a statistical point of view. Named conflicts are in fact artificial tags created by historians to aggregate events that share important historical, geographical and political characteristics, but that may have never really existed as a single event. Under the portmanteau Mongolian Invasions (or Conquests), historians collect all conflicts related to the expansion of the Mongol empire during the thirteenth and fourteenth centuries. Another example is the so-called Hundred-Years' War in the period 1337-1453.  Aggregating all these events necessarily brings to the creation of very large fictitious conflicts accounting for hundreds of thousands or million casualties. The fact that, for historical and historiographical reasons, these events tends to be more present in antiquity and the Middle Ages could bring to a naive overestimation of the severity of wars in the past. Notice that named conflicts like the Mongolian Invasions are different from those like WW1 or WW2, which naturally also involved several tens of battles in very different locations, but which took place in a much shorter time period, with no major time separation among conflicts.

A straightforward technique could be to set a cutoff-point of 25 years for any single event --an 80 years event would be divided into three minor ones. However it remains that the length of the window remains arbitrary. Why 25 years? Why not 17?

As we show in Section \ref{eanalysis}, a solution to all these problems with data is to consider each single observation as an imprecise estimate, a fuzzy number in the definition of \cite{Viertl}. Using Monte Carlo methods we have shown that, if we assume that the real number of casualties in a conflict is uniformly distributed between the minimum and the maximum in the available data, the tail exponent $\xi$ is not really affected (apart from the obvious differences in the smaller decimals).\footnote{Our results do not depend on choice of the uniform distribution, selected for simplification. All other bounded distributions produce the same results in the limit.} Similarly, our results remain invariant if we remove/add a proportion of the observations by bootstrapping our sample and generating new data sets.

We believe that this approach to data as imprecise observations is one of the novelties of our work, which makes our conclusions more robust to scrutiny about the quality of data. We refer to Section \ref{eanalysis} for more details.

\subsection{Missing events} \label{miss}
We are confident that there are many conflicts that are not part of our sample. For example we miss all conflicts among native populations in the American continent before its European discovery, as no source of information is actually available. Similarly we may miss some conflicts of antiquity in Europe, or in China in the sixteenth century. However, we can assume that the great part of these conflicts is not in the very tail of the distribution of casualties, say in the top 10 or 20\%. It is in fact not really plausible to assume that historians have not reported a conflict with 1 or 2 million casualties, or that such an event is not present in \cite{Necro} and \cite{Phillips}, at least for what concerns Europe, Asia and Africa. 

Nevertheless, in Subsection \ref{missingdata}, we deal with the problem of missing data in more details.

\subsection{The conflict generator process}
We are not assuming the existence of a unique conflict generator process. 
It is clear that all conflicts of humanity do not share the same set of causes. Conflicts belonging to different centuries and continents are likely to be not only independent, as already underlined in \cite{Richardson} and \cite{Hayek}, but also to have different origins. 
We thus avoid performing time series analysis beyond the straightforward investigation of the existence of trends. While it could make sense to subject the data to specific tests, such as whether an increase/decrease in war casualties is caused by the  lethality of weapons, it would be unrigorous and unjustified to look for an autoregressive component. How could the An Lushan rebellion in China (755 AD) depend on the Siege of Constantinople by Arabs (717 AD), or affect the Viking Raids in Ireland (from 795 AD on)?  But this does not mean that all conflicts are independent: during WW2, the attack on Pearl Harbor and the Battle of France were not independent, despite the time and spatial separation, and that's why historians merge them into one single event. And while we can accept that most of the causes of WW2 are related to WW1, when studying numerical casualties, we avoid translating the dependence into the magnitude of the events: it would be naive to believe that the number of victims in 1944 depended on the death toll in 1917. How could the magnitude of WW2 depend on WW1?

Critically, when related conflicts are aggregated, as in the case of WW2 or for the Hundred Years' war (1337-1453), our data show that the number of conflicts over time ---  if we focused on very destructive events in the last 500/600 years and we believed in the existence of a conflict generator process --- is likely to follow a homogeneous Poisson process, as already observed by \cite{Hayek} and \cite{Richardson}, thus supporting the idea that wars are randomly distributed accidents over time, not following any particular trend. We refer to Section \ref{eanalysis} for more details.

\subsection{Data transformations}\label{rescaling}
We used three different types of data in our analyses: raw data, rescaled data and dual data.

\subsubsection{Raw data} Presented as collected from the different sources, and shown in Table \ref{data}. Let $X_t$ be the number of casualties in a given conflict at time $t$, define the triplet $\{X_t, X_t^l, X_t^u\ \}$, where $X_t^l$ and $X_t^u$ represent the lower and upper bound, if available.

\subsubsection{Rescaled data} The data is rescaled with respect to the current world population (7.2 billion\footnote{According to the United Nations Department of Economic and Social Affairs \cite{UN}}). Define the triplet $\{Y_t=X_t\frac{P_{2015}}{P_t}, Y_t^l=X_t^l\frac{P_{2015}}{P_t}, Y_t^u=X_t^u\frac{P_{2015}}{P_t}\ \}$, where $P_{2015}$ is the world population in 2015 and $P_t$ is the population at time $t=1,...,2015$.\footnote{If for a given year, say 788 AD, we do not have a population estimate, we use the closest one.} Rescaled data was used by Steven Pinker \cite{Pinker} to account for the relative impact of a conflict. 
This rescaling tends to inflate past conflicts, and can lead to the naive statement that violence, if defined in terms of casualties, has declined over time. But we agree with those scholars like Robert Epstein \cite{Epstein} stating  ``[...] why should we be content with only a relative decrease? By this logic, when we reach a world population of nine billion in 2050, Pinker will conceivably be satisfied if a mere two million people are killed in war that year". 
In this paper we used rescaled data for comparability reasons, and to show that even with rescaled data we cannot really state statistically that war casualties have declined over time (and thus violence).

\subsubsection{Dual data} those we obtain using the log-rescaling of equation (\ref{transform}), the triplet $\{Z_t=\varphi(Y_t),Z_t^l=\varphi(Y_t^l),Z_t^u=\varphi(Y_t^u)\}$. The input for the function $\varphi(\cdot)$ are the rescaled values $Y_t$.

Removing the upper bound allows us to combine practical convenience with theoretical rigor. Since we want to apply the tools of extreme value theory, the finiteness of the upper bound is fundamental to decide whether a heavy-tailed distribution falls in the Weibull (finite upper bound) or in the Fr\'echet class (infinite upper bound). Given their finite upper bound, war casualties are necessarily in the Weibull class. From a theoretical point of view the difference is large \cite{deHaan}, especially for the existence of moments. But from a practical point of view, heavy-tailed phenomena are better modeled within the Fr\'echet class. Power laws belong to this class. And as observed by \cite{Embrechts}, when a fat-tailed random variable $Y$ has a very large finite upper bound $H$, unlikely to be ever touched (and thus observed in the data), its tail, which from a theoretical point of view falls in the so-called Weibull class, can be modeled as if belonging to the Fr\'echet one. In other words, from a practical point of view, when $H$ is very large, it is impossible to distinguish the two classes; and in most of the cases this is indeed not a problem.

Problems arise when the tail looks so thick that even the first moment appears to be infinite, when we know that this is not possible. And this is exactly the case with out data, as we show in the next Section.

\begin{figure}[!htb]
\includegraphics[width=\linewidth]{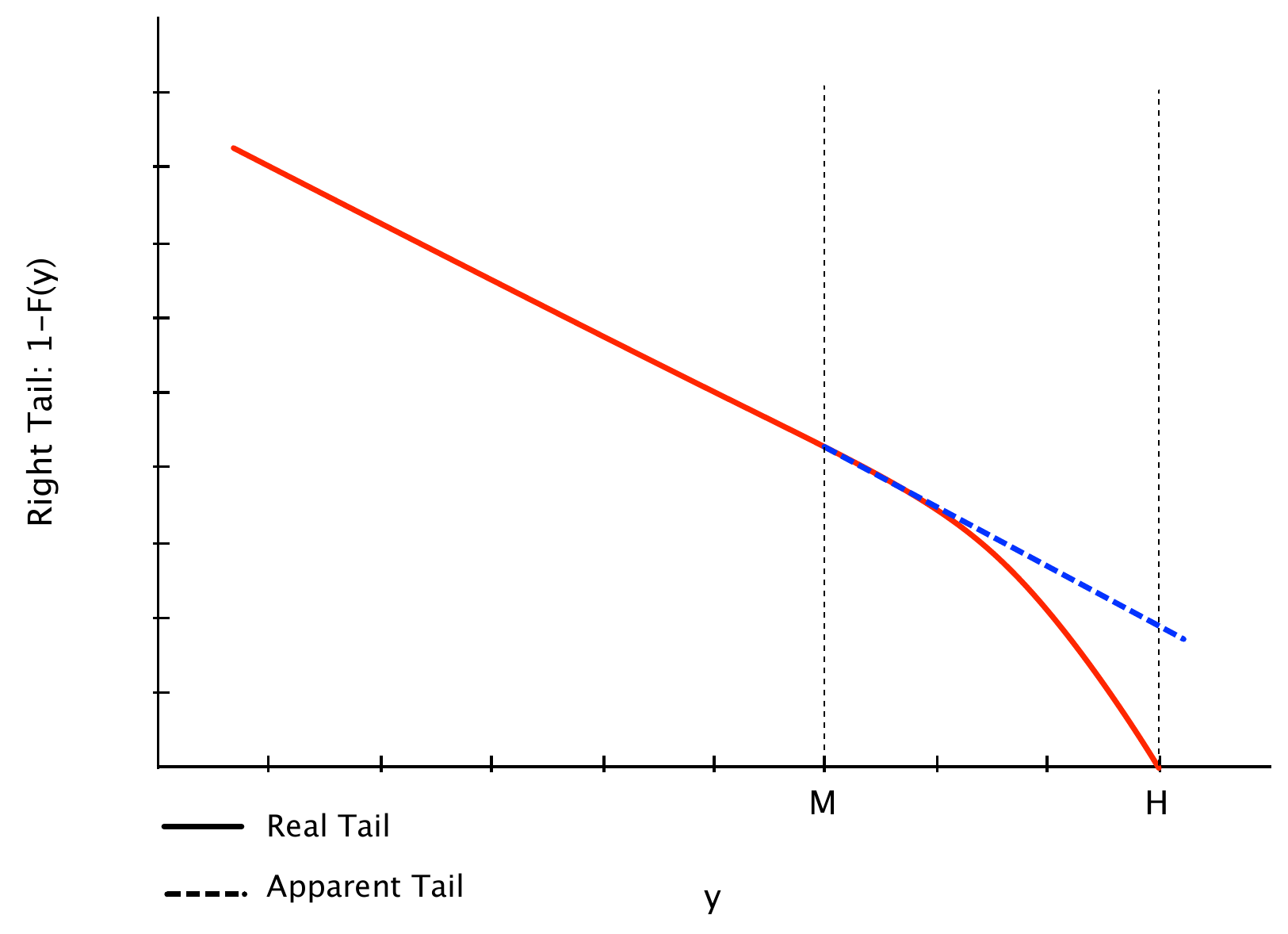}
\caption{Graphical representation (Log-Log Plot) of what may happen if one ignores the existence of the finite upper bound $H$, since only $M$ is observed.}
\label{tailcomp}
\end{figure}

Figure \ref{tailcomp} shows illustrates the need to separate real and dual data. For a random variable $Y$ with remote upper bound $H$, the real tail is represented by the continuous line. However, if we only observe values up to $M$, and we ignore the existence of $H$, which is unlikely to be reached and therefore hardly observable, we could be inclined to believe the the tail is the dotted one, the apparent one. The two tails are indeed essentially indistinguishable for most cases, but the divergence is evident when we approach $H$.  Hence our transformation. Notice that $\varphi(y)\approx y$ for very large values of $H$. This means that for a very large upper bound, unlikely to be reached as the world population, the results we get for the tail of $Y$ and $Z=\varphi(Y)$ are essentially the same most of the times. This is exactly what happens with our data, considered that no armed conflict has ever killed more than 19\% of the world population (the Three Kingdoms, 184-280 AD). But while $Y$ is bounded, $Z$ is not. 

Therefore we can safely model $Z$ as belonging to the Fr\'echet class, study its tail behavior, and then come back to $Y$ when we are interested in the first moments, which under $Z$ could not exist.

\section{Descriptive Data Analysis} \label{danalysis}
Figure \ref{bubbles} showing casualties over time, is composed of four subfigures, two related to raw data and two related to dual data\footnote{Given Equation (\ref{transform}), rescaled and dual data are approximately the same, hence there is no need to show further pictures.}. For each type of data, using the radius of the different bubbles, we show the relative size of each event in terms of victims with respect to the world population and within our data set (what we call Impact). Note that the choice of the type of data (or of the definition of size) may lead to different interpretation of trends and patterns. From rescaled and dual data, one could be lead to superficially infer a decrease in the number of casualties over time. 

As to the number of armed conflicts, Figure \ref{bubbles} seems to suggest an increase over time, as we see that most events are concentrated in the last 500 years or so, an apparent illusionary increase most likely due to a reporting bias for the conflicts of antiquity and early Middle Ages.

\begin{figure}[!htb]
\includegraphics[width=\linewidth]{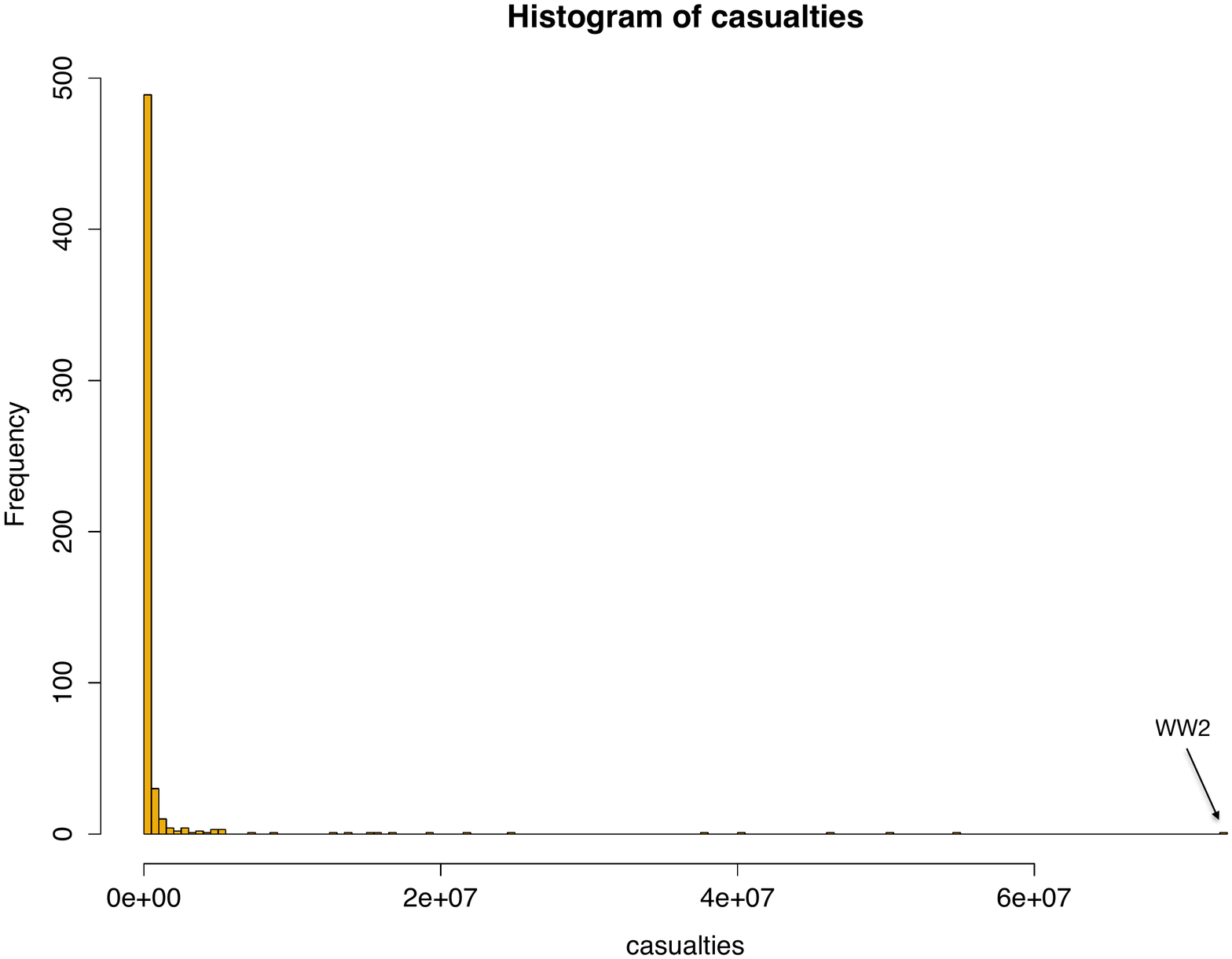}
\caption{Histogram of war casualties, using raw data. A long fat right tail is clearly visible.}
\label{Hist}
\end{figure}

Figure \ref{Hist} shows the histogram of war casualties, when dealing with raw data. Similar results are obtained using rescaled and dual casualties. The graphs suggests the presence of a long right tail in the distribution of victims, and the maximum is represented by WW2, with an estimated amount of about 73 million victims.

\begin{figure}[!htb]
\includegraphics[width=\linewidth]{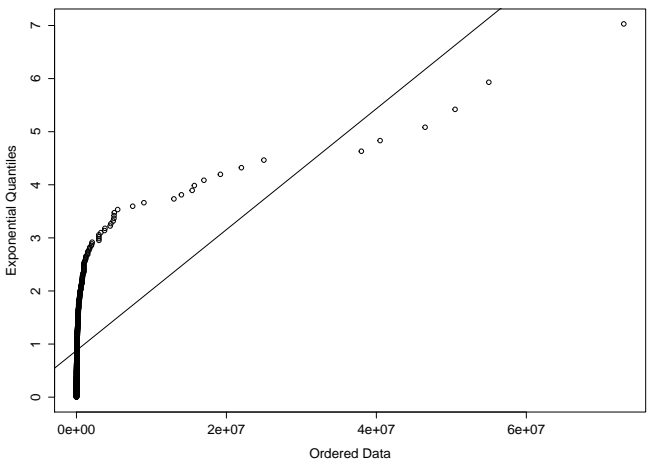}
\caption{Exponential qq-plot of war casualties, using raw data. The clear concave behavior suggest the presence of a right fat tail.}
\label{QQplot}
\end{figure}

The presence of a Paretian tail is also supported by the quantile-quantile plot in Figure \ref{QQplot}, where we use the exponential distribution as a benchmark. The concave behavior we can observe in the plot is considered a good signal of fat-tailed distribution \cite{Embrechts}, \cite{Klei}.

In order to investigate the presence of a right fat tail, we can make use of another graphical tool: the \textit{meplot}, that is, mean excess function plot.

Let $X$ be a random variable with distribution $F$ and right endpoint $x_F$ (i.e. $x_F=\sup\{x \in \mathbb{R}: F(x)<1\}$). The function
\begin{equation}
e(u)=E[X-u|X>u]=\frac{\int_u^\infty (t-u) \text{d}F(t)}{\int_u^\infty \text{d}F(t)}, \qquad 0<u<x_F,
\end{equation}
is called mean excess function of $X$ (mef). The empirical mef of a sample $X_{1}$, $X_{2}$,..., $X_{n}$ is easily computed as
\begin{equation}
e_{n}(u)=\frac{\sum_{i=1}^{n}(X_{i}-u)}{\sum_{i=1}^{n}1_{\{X_{i}>u\}}},
\end{equation}
that is the sum of the exceedances over the threshold $u$ divided by the number of such data points. Interestingly, the mean excess function is a way of characterizing distributions within the class of continuous distributions \cite{Klei}. For example, power law distributions are characterized by the van der Wijk's law \cite{vanWijk}, that is, by a mean excess function linearly increasing in the threshold $u$, while an exponential distribution of parameter $\lambda$ would show a constant mean excess function with value $\lambda$. 

\begin{figure}[!htb]
\includegraphics[width=\linewidth]{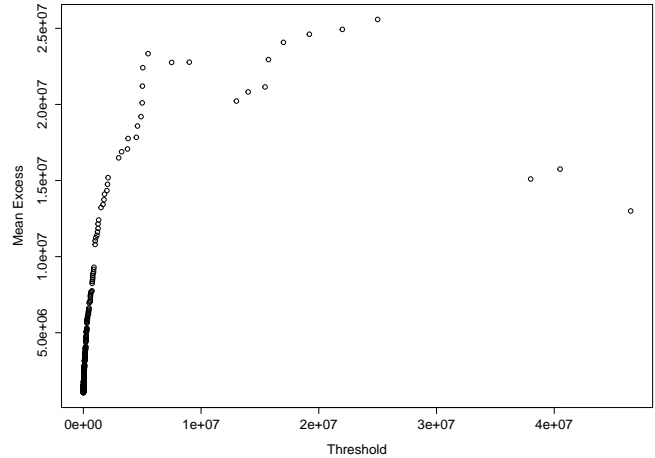}
\caption{Mean Excess Function Plot (meplot) for war casualties, using raw data. A steep upward trend is visible, suggesting the presence of a Paretian right tail.}
\label{Meplot}
\end{figure}

 Figure \ref{Meplot} shows the the meplot of war casualties, again using raw data. The graph is easily obtained by plotting the pairs $\{(X_{i:n},e_n(X_{i:n})): i=1,...,n\}$, where $X_{i:n}$ is the $i-$th order statistic, used as a threshold. The upward trend in Figure \ref{Meplot} is a further signal of fat-tailed data, as discussed for example in \cite{Cirillo} and \cite{Embrechts}.

A characteristic of the power law class is the non-existence (infiniteness) of moments, at least the higher-order ones, with important consequences in terms of statistical inference. If the variance is not finite, as is often the case with financial data, it becomes more problematic to build confidence intervals for the mean. Similarly, if the third moment (skewness) is not defined, it is risky to build confidence intervals for the variance.

An interesting graphical tool showing the behavior of moments in a given sample, is the \textit{Maximum-to-Sum plot}, or MS Plot. The MS Plot relies on simple consequence of the law of large numbers \cite{Embrechts}. For a sequence $X_1,X_2,...,X_n$ of nonnegative i.i.d. random variables, if for $p=1,2,3...$, $E[X^p]<\infty$, then $R_n^p=M_n^p / S_n^p \to^{a.s.} 0$ as $n \to \infty$, where $S_n^p=\sum_{i=1}^n X_i^p$ is the partial sum, and $M_n^p=\max(X_1^p,...,X_n^p)$ the partial maximum.

\begin{figure}[!htb]
\includegraphics[width=\linewidth]{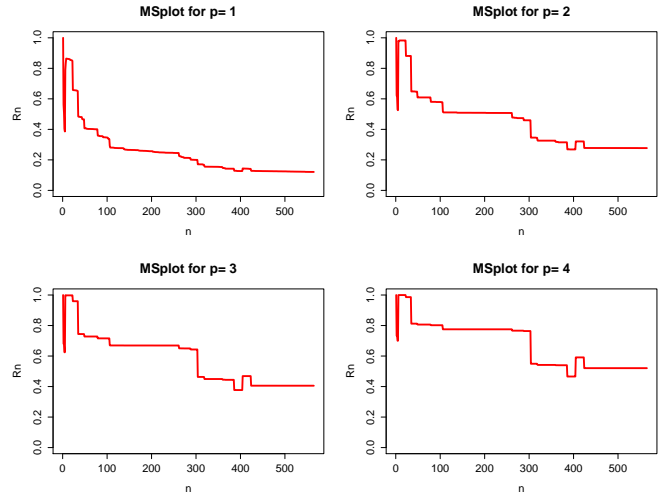}
\caption{Maximum to Sum plot to verify the existence of the first four moments of the distribution of war casualties, using raw data. No clear converge to zero is observed, for $p=2,3,4$, suggesting that the moments of order 2 or higher may not be finite.}
\label{MSPlot_raw}
\end{figure}

\begin{figure}[!htb]
\includegraphics[width=\linewidth]{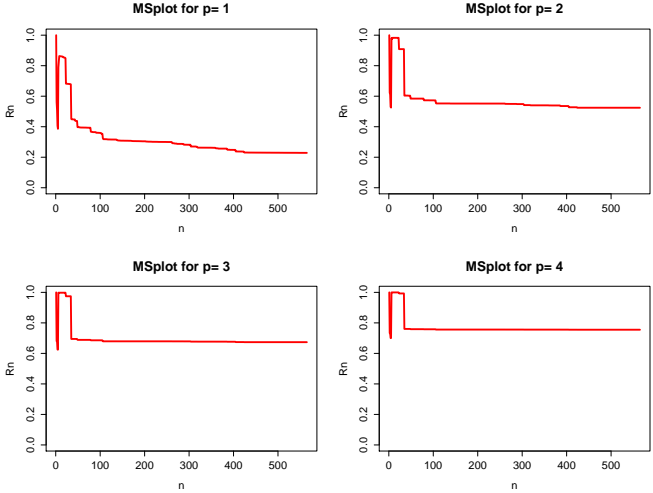}
\caption{Maximum to Sum plot to verify the existence of the first four moments of the distribution of war casualties, using rescaled data. No clear converge to zero is observed, suggesting that all the first four moments may not be finite.}
\label{MSPlot_pink}
\end{figure}

Figures \ref{MSPlot_raw} and \ref{MSplot_pink} show the MS Plots of war casualties for raw and rescaled data respectively. No finite moment appears to exist, no matter how much data is used: it is in fact clear that the $R_n$ ratio does not converge to 0 for $p=1,2,3,4$, thus suggesting that the distribution of war casualties has such a fat right tail that not even the first moment exists. This is particularly evident for rescaled data. 

\begin{figure}[!htb]
\includegraphics[width=\linewidth]{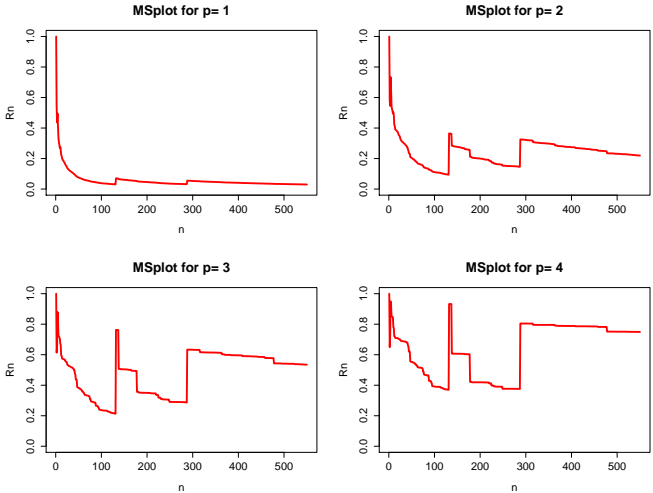}
\caption{Maximum to Sum plot of a Pareto(1.5): as expected, the first moment is finite ($R_n\to0$), while higher moments do not exists ($R_n^p$ is erratic and bounded from 0 for $p=2,3,4$).}
\label{MSCompare}
\end{figure}

To compare to thin tailed situations, \ref{MSCompare} shows the MS Plot of a Pareto $(\alpha=1.5)$: notice that the first moment exists, while the other moments are not defined, as the shape parameter equals $1.5$. For thin tailed distributions such as the Normal, the MS Plot rapidly converges to 0 for all $p$.

\begin{table}
\caption{Average inter-arrival times (in years) and their mean absolute deviation for events with more than 0.5, 1, 2, 5, 10, 20 and 50 million casualties, using raw (Ra) and rescaled data (Re).}
\begin{center}
\begin{tabular}{|c|c|c|c|c|}
\hline
\textbf{Thresh.} &\textbf{Avg Ra}& \textbf{MAD Ra} &\textbf{Avg Re} & \textbf{MAD Re}\\
\hline
0.5 & 23.71& 35.20 & 9.63& 15.91 \\
1 & 34.08 & 47.73 & 13.28 & 20.53\\
2 & 56.44 & 72.84 & 20.20  & 28.65 \\
5 & 93.03& 113.57 & 34.88 &46.85 \\
10 & 133.08 & 136.88  & 52.23 & 63.91\\
20 & 247.14 & 261.31 &73.12  & 86.19\\
50 & 345.50 & 325.50 &103.97  & 114.25\\
\hline
\end{tabular}
\end{center}
\label{times}
\end{table}%

Table \ref{times} provides  information about armed conflicts and their occurrence over time. 
Note that we have chosen very large thresholds to minimize the risk of under-reporting.  We  compute the distance, in terms of years, between two time-contiguous conflicts, and use for measure of dispersion the mean absolute deviation (from the mean). Table \ref{times} shows the average inter-arrival times between armed conflicts with at least 0.5, 1, 2, 5, 10, 20 and 50 million casualties. 
For a conflict of at least 500k casualties, we need to wait on average 23.71 years using raw data, and 9.63 years using rescaled or dual data (which, as we saw, tend to inflate the events of antiquity). For conflicts with at least 5 million casualties, the time delay is on average 93.03 or 34.88, depending on rescaling. 
Clearly, the bloodier the conflict, the longer the inter-arrival time. The results essentially do not change if we use the mid or the ending year of armed conflicts.

The consequence of this analysis is that the absence of a conflict generating more than, say, 5 million casualties in the last sixty years highly insufficient to state that their probability has decreased over time, given that the average inter-arrival time is 93.03 years, with a mean absolute deviation of 113.57 years! 
Unfortunately, we need to wait for more time to assert whether we are really living in a more peaceful era,: present data are not in favor (nor against) a structural change in violence, when we deal with war casualties.

\begin{figure}[!htb]
\includegraphics[width=\linewidth]{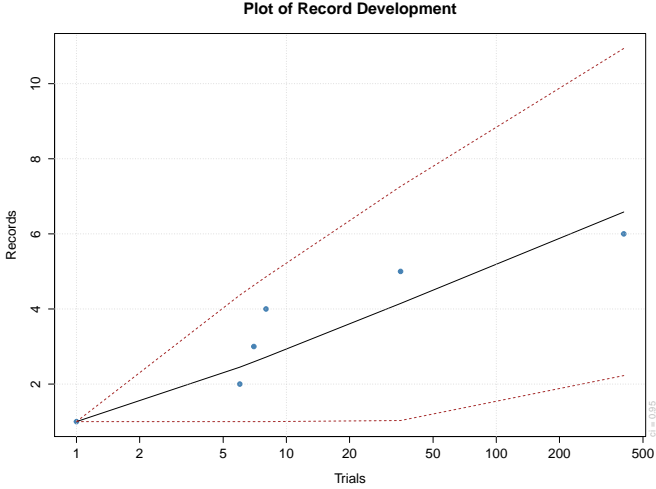}
\caption{Record plot for testing the plausibility of the i.i.d. hypothesis, which seems to be supported.}
\label{Records}
\end{figure}

Section \ref{data} asserted that our data set do not form a proper time series, in the sense that no real time dependence is present (apart from minor local exceptions).\footnote{Even if redundant, given the nature of our data and the descriptive analysis performed so far, the absence of a relevant time dependence can be checked by performing a time series analysis of war casualties. No significant trend, season or temporal dependence can be observed over the entire time window.} Such consideration is further supported by the so-called record plot in Figure \ref{Records}. This graph is used to check the i.i.d. nature of data, and it relies on the fact that, if data were i.i.d., then records over time should follow a logarithmic pattern \cite{Embrechts}. Given a sequence $X_1,X_2,...$, and defined the partial maximum $M_n$, an observation $x_n$ is called a record if $X_n=M_n=x_n$.

For the sake of completeness, if we focus our attention on shorter periods, like the 50 years following WW2, a reduction in the number and the size of the conflict can probably be observed --- the so-called Long Peace of \cite{Goldstein}, \cite{Mueller} and \cite{Pinker}; but recent studies suggest that this trend could already have changed in the last years \cite{Inkster}, \cite{Norton}. However, in our view, this type of analysis is not meaningful, once we account for the extreme nature of armed conflicts, and the long inter-arrival times between them, as shown in Section \ref{danalysis}. 

\section{Tail risk of armed conflicts} \label{eanalysis}
Our data analysis in Section \ref{danalysis} suggesting a heavy right tail for the distribution of war casualties, both for raw and rescaled data is consistent with the existing literature, e.g. \cite{Clauset}, \cite{Friedman}, \cite{Scharpf} and their references.
Using extreme value theory, or EVT \cite{Balkema78}, \cite{Balkema79}, \cite{deHaan}, \cite{Embrechts}, \cite{Gumbel}, \cite{Taleb}, we use the Generalized Pareto Distribution (GPD). The choice is due to the Pickands, Balkema and de Haan's Theorem \cite{Balkema74}, \cite{Pickands}.

Consider a random variable $X$ with distribution function $F$, and call $F_u$ the conditional df of $X$ above a given threshold $u$. We can then define the r.v. $W$, representing the rescaled excesses of $X$ over the threshold $u$, so that
$$F_u(w) = P(X-u \leq w | X>u) = \frac{F(u+w)-F(u)}{1-F(u)}, $$
for $0 \leq w \leq x_F-u$, where $x_F$ is the right endpoint of the underlying distribution $F$. 

Pickands, Balkema and de Haan \cite{Balkema74}, \cite{Pickands} have showed that for a large class of underlying distribution functions $F$, and a large $u$, $F_u$ can be approximated by  a Generalized Pareto distribution (hence GPD), i.e.
$F_u(w) \rightarrow G(w),\text{ as }u \rightarrow \infty$
where
\begin{equation}
G (w)=\begin{cases}
1-(1+\xi \frac{w}{\sigma})^{-1/\xi} &  if \,\xi \neq 0\\
 1-e^{-\frac{w}{\sigma}} &  if \,\xi = 0
\end{cases}, \; w\geq 0.
\label{GPD}
\end{equation}
The parameter $\xi$, known as the shape parameter, is the central piece of information for our analysis: $\xi$ governs the fatness of the tails, and thus the existence of moments. The moment of order $p$ of a Generalized Pareto distributed random variable only exists if and only if $\xi < 1/p$ \cite{Embrechts}.

The Pickands, Balkema and de Haan's Theorem thus allows us to focus on the right tail of the distribution without caring too much about what happens below the threshold $u$. 
One powerful property of the generalized Pareto is the tail stability with respect to threshold \cite{Embrechts}. Formally, if $W\sim \text{GPD}(\xi,\sigma_1)$, for $W\geq u_1$, then $W\sim \text{GPD}(\xi,\sigma_2)$ for $W\geq u_2>u_1$. In other words: increasing the threshold does not affect the shape parameter. What changes is only the scale parameter.

We start by identifying the threshold $u$ above which the GPD approximation may hold. Different heuristic tools can be used for this purpose, from Zipf plots, i.e. log log plots of the survival function, to mean excess function plots (as the one in Figure \ref{Meplot}), where one looks for the threshold above which it is possible to visualize --- if present --- the linearity that is typical of fat-tailed phenomena. Other possibilities like the Hill plot are discussed for example in \cite{deHaan}.

Our investigations suggest that the best threshold for the fitness of GPD is $25,000$  casualties, when using raw data, well above the $3,000$ minimum we have set in collecting our observations. A total of 331 armed conflicts lie above this threshold (58.6\% of all our data).

The use of rescaled and dual data requires us to rescale the threshold; the value thus becomes 145k casualties. It is worth noticing that, nevertheless, for rescaled data the threshold could be lowered to 50k, and still we would have a satisfactory GPD fitting.

If $\xi>-1/2$, the parameters of a GPD can be easily estimated using Maximum Likelihood Estimation (MLE) \cite{deHaan}, while for $\xi \leq -1/2$ MLE estimates are not consistent and other methods are better used \cite{Embrechts}.

\begin{figure}[!htb]
\includegraphics[width=\linewidth]{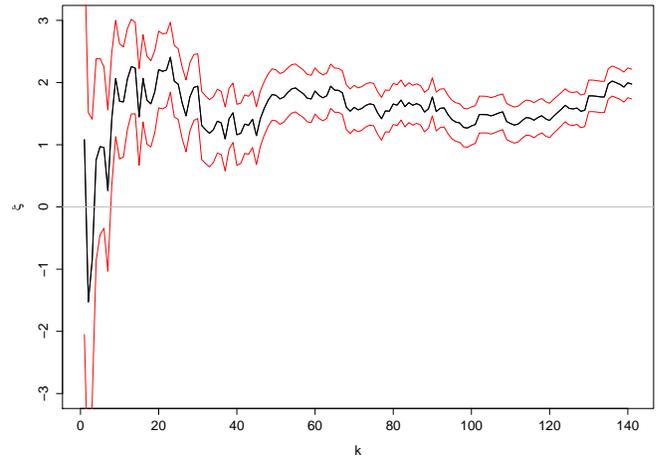}
\caption{Pickands' estimator for the $\xi$ shape parameter for war casualties (raw data). The value 1.5 appears to be a good educated guess.}
\label{Pickands}
\end{figure}

In our case, given the fat-tailed behavior, $\xi$ is likely to be positive. This is clearly visible in Figure \ref{Pickands}, showing the Pickands plot, based on the Pickands' estimator for $\xi$, defined as 
$$
\hat{\xi}^{(P)}_{k,n}=\frac{1}{\log 2} \log \frac{X_{k,n}-X_{2k,n}}{X_{2k,n}-X_{4k,n}},
$$
where $X_{k,n}$ is the $k$-th upper order statistics out of a sample of $n$ observations.

\begin{table}[!htb]
\caption{Maximum likelihood estimates (and standard errors) of the Generalized Pareto Distribution parameters for casualties over different thresholds, using raw, rescaled and dual data. We also provide the number of events lying above the thresholds, the total number of observations being equal to 565.}
\begin{center}
\begin{tabular}{|l|c|c|c|}
\hline
\textbf{Data} &Threshold &$\boldsymbol\xi$ & $\boldsymbol\sigma$\\
\hline
Raw Data & 25k& 1.4985 & 90620 \\
 &    &\small{(0.1233)}&  \small{(2016)} \\
 \hline
Rescaled Data &145k & 1.5868& 497436 \\
 &  &\small{(0.1265)}& \small{(2097)} \\
 \hline
Dual Data  &145k & 1.5915& 496668 \\
 &  &\small{(0.1268)}& \small{(1483)} \\
\hline
\end{tabular}
\end{center}
\label{estimates}
\end{table}%

In the Pickands plot, $\hat{\xi}^{(P)}_{k,n}$ is computed for different values of $k$, and the "optimal" estimate of $\xi$ is obtained by looking at a more or less stable value for $\hat{\xi}^{(P)}_{k,n}$ in the plot. In Figure \ref{Pickands}, the value $1.5$ seems to be a good educated guess for raw data, and similar results hold for rescaled and dual amounts. In any case, the important message is that $\xi>0$, therefore we can safely use MLE.

Table \ref{estimates} presents our estimates of $\xi$ and $\sigma$ for raw, rescaled and dual data. In all cases, $\xi$ is significantly greater than 1, and around 1.5 as we guessed by looking at Figure \ref{Pickands}. This means that in all cases the mean of the distribution of casualties seems to be infinite, consistently with the descriptive analysis of Section \ref{danalysis}.

Further, Table \ref{estimates} shows the similarity between rescaled and dual data. Looking at the standard errors, no test would reject the null hypothesis that $\xi_{rescaled}=\xi_{dual}$, or that $\sigma_{rescaled}=\sigma_{dual}$. 

\begin{figure}[!htb]
\begin{subfigure}{.5\textwidth}
\includegraphics[width=\linewidth]{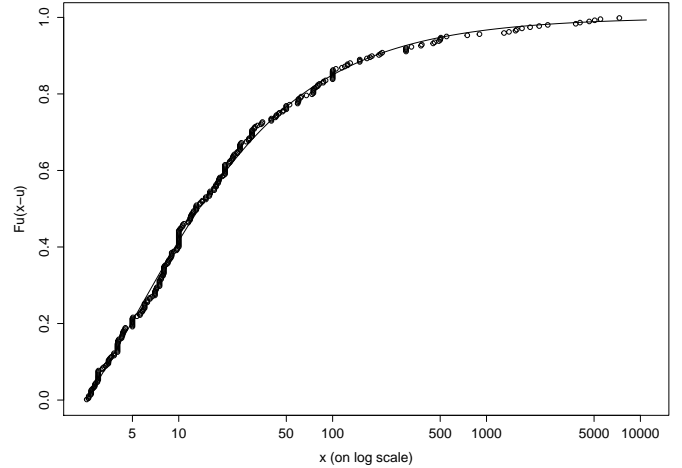}
\caption{Distribution of the excesses (ecfd and theoretical).}
\label{exdf}
\end{subfigure}
\begin{subfigure}{.5\textwidth}
\includegraphics[width=\linewidth]{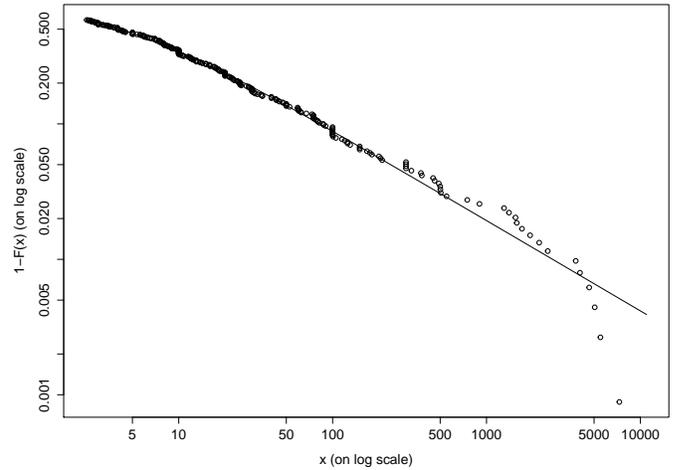}
\caption{Right Tail (empirical and theoretical).}
\label{extail}
\end{subfigure}
\caption{GPD fitting of war casualties exceeding the 145k threshold, using dual data.}
\label{Fitting}
\end{figure}

\begin{figure}[!htb]
\begin{subfigure}{.5\textwidth}
\includegraphics[width=\linewidth]{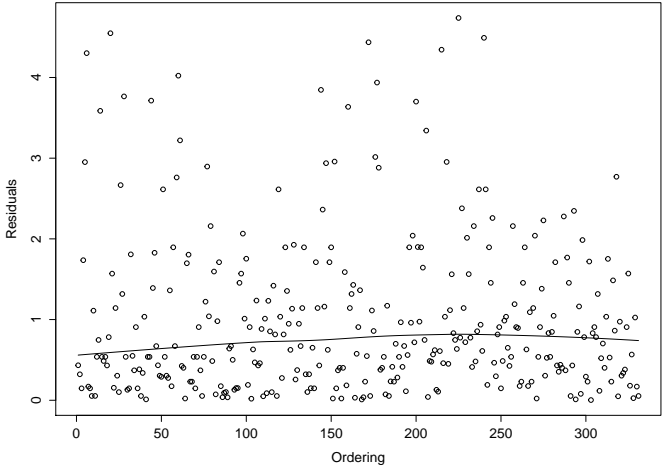}
\caption{Scatter plot of residuals and their smoothed average. }
\label{exdf}
\end{subfigure}
\begin{subfigure}{.5\textwidth}
\includegraphics[width=\linewidth]{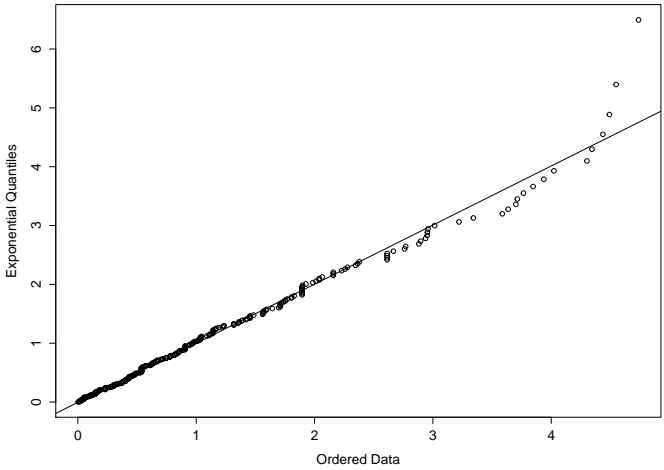}
\caption{QQ-plot of residuals against exponential quantiles.}
\label{extail}
\end{subfigure}
\caption{Residuals of the GPD fitting: looking for exponentiality.}
\label{Residuals}
\end{figure}

Figure \ref{Fitting} shows the goodness of our GPD fit for dual data, which is also supported by goodness-of-fit tests for the Generalized Pareto Distribution \cite{Villa} (p-value: 0.37). As usual, similar results do hold for raw and rescaled data. Figure \ref{Residuals} shows the residuals of the GPD fit which are, as can be expected (see \cite{Embrechts}), exponentially distributed.


\section{Estimating the shadow mean} \label{ghost}
Given the finite upper bound $H$, we known that $E[Y]$ must be finite as well. A simple idea is to estimate it by computing the sample mean, or the conditional mean above a given threshold\footnote{Notice that the sample mean above a minimum threshold $L$ corresponds to the concept of expected shortfall used in risk management.}, if we are more interested in the tail behavior of war casualties, as in this paper. For a minimum threshold $L$ equal to 145k, this value is $1.77 \times 10^7$ (remember that $Y$ represents the rescaled data). 

However, in spite of its boundedness, the support of $Y$ is so large that $Y$ behaves like a very heavy-tailed distribution, at least until we do not approach the upper bound $H$. This makes the (conditional) sample mean not really reliable, or robust: one single extreme observation can make it jump, given that the (conditional) sample mean has a breakdown point equal to 0 \cite{Maronna}.

A more robust way of computing what we define the conditional shadow mean of $Y$ (the true conditional mean not visible from data) is therefore to use the log-transformation of Equation (\ref{transform}).

Let $F$ and $G$ be respectively the distribution of $Y$ and $Z$. With $f$ and $g$, we indicate the densities.

In the previous section we have seen that, when using the dual version of rescaled data, $G\approx \text{GPD}(\xi,\sigma)$ for $Z\geq L=145000$ (even 50k gives good results, but 145k is comparable to the 25k threshold of raw data). Moreover we know that $Z=\varphi(Y)$, so that $Y=\varphi^{-1}(Z)=(L-H) e^{\frac{L-Z}{H}}+H$. 

This implies that the distribution of $Y$, for $Y\geq L$, can be obtained from the distribution of $G$.

First we have
\begin{equation}
\int_L^\infty g(z) \, \mathrm{d} z = \int_L^{\varphi^{-1}(\infty)} f(y) \, \mathrm{d} y.
\end{equation}
And we know that
\begin{equation}
g(z;\xi,\sigma)=\frac{1}{\sigma} \left(1+\frac{\xi z}{\sigma}\right)^{-\frac{1}{\xi}-1},\qquad z\in [L,\infty).
\end{equation}
This implies, setting $\alpha=\xi^{-1}$ and $k=\sigma/H$, 
\begin{equation}
f(y;\alpha,k)=\frac{\left(1- \frac{\log(H-y)-\log(H-L)}{\alpha k} \right)^{-\alpha-1}}{(H-y)k}, \qquad y \in [L,H].
\end{equation}
We can then derive the shadow mean of $Y$ as
\begin{equation}
E[Y]=\int_L^H y \, f(y) \,\mathrm{d} y ,
\end{equation}
obtaining
\begin{equation}
E[Y]= L+(H-L)e^{\alpha k} (\alpha k)^\alpha \Gamma\left(1-\alpha, \alpha k \right). 	
\end{equation}

Table \ref{means} provides the shadow mean, the sample mean and their ratio for different minimum thresholds $L$, using rescaled data. As already said, our GPD fit is already satisfactory for $L=50000$, that is why we start from that threshold. 

The ratios in Table \ref{means} show how the sample mean underestimates the shadow mean, especially for lower thresholds, for which the shadow mean is almost 3 times larger than the sample mean. For these reasons, a "journalistic" reliance on sample mean would be highly misleading, when considering all conflict together, without setting large thresholds. There would be the serious risk of underestimating the real lethality of armed conflicts.

\begin{table}[h!]
\caption{Conditional shadow mean, conditional sample mean and their ratio for different minimum thresholds. In bold the values for the 145k threshold. Rescaled data.}
\begin{center}
\begin{tabular}{|c|c|c|c|}
\hline
\textbf{Thresh.$\times 10^3$}  &\textbf{Shadow$\times 10^7$}& \textbf{Sample$\times 10^7$}& \textbf{Ratio}\\
\hline
 50 & 3.6790 &  1.2753 & 2.88 \\
 100 &  3.6840 &  1.5171 & 2.43 \\
 \textbf{145} & \textbf{3.6885} & \textbf{1.7710}& \textbf{2.08} \\
 300 & 4.4089& 2.2639& 1.95\\
 500 & 5.6542 &  2.8776 & 1.96 \\
1000 & 6.5511 &  3.9481 &  1.85 \\
 \hline
\end{tabular}
\end{center}
\label{means}
\end{table}%


A similar procedure can be applied to raw data. Fixing $H$ equal to 7.2 billion and $L=25000$, we can define $V=\varphi(X)$ and study its tail. The tail coefficient $\xi_V$ is $1.4970$. Even in this case the sample mean underestimates the shadow mean by a factor of at least 1.6 (up to a maximum of 7.5!).

These results tell us something already underlined by \cite{Epstein}: we should not underestimate the risk associated to armed conflicts, especially today. While data do not support any significant reduction in human belligerence, we now have both the connectivity and the technologies to annihilate the entire world population (finally touching that value $H$ we have not  touched so far). In the antiquity, it was highly improbable for a conflict involving the Romans to spread all over the world, thus also affecting the populations of the Americas. A conflict in ancient Italy could surely influence the populations in Gaul, but it could have no impact on people living in Australia at that time. Now, a conflict in the Middle East could in principle start the next World War, and no one could feel completely safe.

It is important to stress that changes in the value of $H$ do not effect the conclusions we can draw from data, when variations do not  modify the order of magnitude of $H$. In other terms, the estimates we get for $H=7.2$ billion are qualitatively consistent to what we get for all values of $H$ in the range $[6,9]$ billion. For example, for $H=9$ billion, the conditional shadow mean of rescaled data, for $L=145000$, becomes $4.0384\times 10^7$. This value is surely larger than $3.6885\times 10^7$, the one we find in Table \ref{means}, but it is still in the vicinity of $4\times 10^7$, so that, from a qualitative point of view, our conclusions are still completely valid.

\subsection*{Other shadow moments}
As we did for the mean, we can use the transformed distribution to compute all moments, given that, since the support of $Y$ is finite, all moments must be finite. However, because of the wide range of variation of $Y$, and its heavy-tailed behavior, one needs to be careful in the interpretation of these moments. As discussed in \cite{Taleb}, the simple standard deviation (or the variance) becomes unreliable when the support of $Y$ allows for extremely fat tails (and one should prefer other measures of variability, like the mean absolute deviation). It goes without saying that for higher moments it is even worse.

Deriving these moments can be cumbersome, both analytically and numerically, but it is nevertheless possible. For example, if we are interested in the standard deviation of $Y$ for $L=145000$, we get $3.08\times 10^8$. This value is very large, three times the sample standard deviation, but finite. Other values are shown in Table \ref{sds}. In all cases, the shadow standard deviation is larger than the sample ones, as expected when dealing with fat-tails \cite{Falk}. As we stressed that care must be used in interpreting these quantities.

Note that our dual approach can be generalized to all those cases in which an event can manifest very extreme values, without going to infinity because of some physical or economical constraint.

\begin{table}[h!]
\caption{Shadow standard deviation, sample standard deviation and their ratio for different minimum thresholds. Rescaled data.}
\begin{center}
\begin{tabular}{|c|c|c|c|}
\hline
\textbf{Thresh.$\times 10^3$}  &\textbf{Shadow$\times 10^8$}& \textbf{Sample$\times 10^8$}& \textbf{Ratio}\\
\hline
 50  &  2.4388 & 0.8574& 2.8444 \\
 100&  2.6934  &  0.9338 & 2.8843 \\
145 & 3.0788 & 1.0120& 3.0422 \\
 300 & 3.3911 & 1.1363 & 2.9843\\
 500  &  3.8848 & 1.2774 & 3.0412 \\
 1000 &  4.6639 & 1.4885  & 3.1332 \\
 \hline
\end{tabular}
\end{center}
\label{sds}
\end{table}%

\section{Dealing with missing and imprecise data} \label{missingdata}
\begin{figure}[!htb]
\begin{subfigure}{.5\textwidth}
\includegraphics[width=\linewidth]{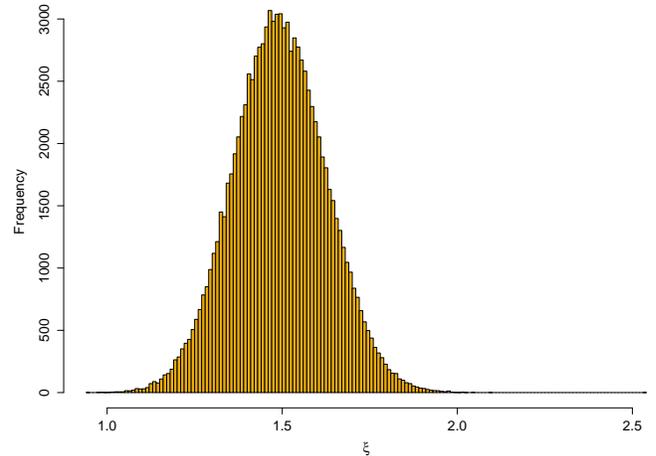}
\caption{Raw data.}
\label{BootRaw}
\end{subfigure}
\begin{subfigure}{.5\textwidth}
\includegraphics[width=\linewidth]{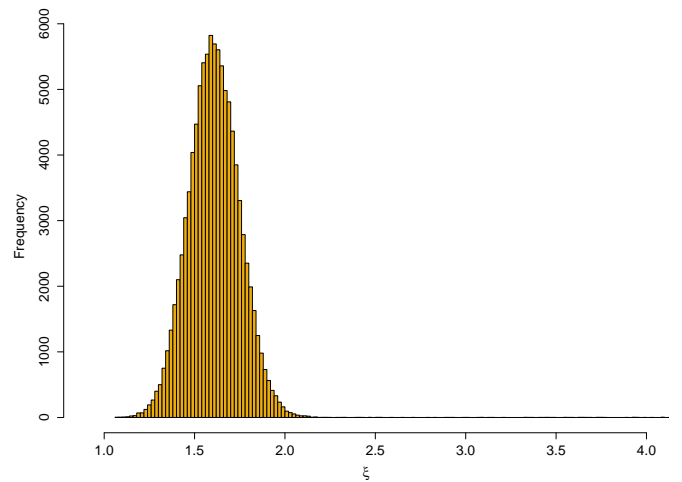}
\caption{Rescaled data.}
\label{BootPink}
\end{subfigure}
\caption{Distribution of the $\xi$ estimate (thresholds 25k and 145k casualties for raw and rescaled data respectively) according to 100k bootstrap samples.}
\label{Boot}
\end{figure}

An effective way of checking the robustness of our estimates to the "quality and reliability" of data is to use resampling techniques, which allow us to deal with non-precise \cite{Viertl} and possibly missing data. We have performed three different experiments:
\begin{itemize}
\item Using the jackknife, we have created 100k samples, by randomly removing up to 10\% of the observations lying above the 25k thresholds. In more than 99\% of cases $\xi>>1$. As can be expected, the shape parameter $\xi$ goes below the value 1 only if most of the observations we remove belong to the upper tail, like WW1 or WW2. 
\item Using the bootstrap with replacement, we have generated another set of 100k samples. As visible in Figure \ref{Boot}, $\xi \leq 1$ in a tiny minority of cases, less than 0.5\%. All other estimates nicely distribute around the values of Table \ref{estimates}. This is true for raw, rescaled and dual data.
\item We tested for the effect of changes in such large events as WW2 owing to the inconsistencies we mentioned earlier (by selecting across the various estimates shown in Table \ref{data}) ---the effect does not go beyond the smaller decimals.
\end{itemize}
We can conclude that our estimates are remarkably robust to imprecisions and missing values in the data\footnote{That's why, when dealing with fat tails, one should always prefer $\xi$ to other other quantities like the sample mean \cite{Taleb}. And also use $\xi$ in corrections, as we propose here.}.

\section{Frequency of armed conflicts}\label{trendarm}
Can we say something about the frequency of armed conflicts over time? Can we observe some trend?
In this section, we show that our data tend to support the findings of \cite{Hayek} and \cite{Richardson}, contra \cite{Mueller}, \cite{Pinker}, that  armed conflict are likely to follow a homogeneous Poisson process, especially if we focus on events with large casualties.

The good GPD approximation allows us to use a well-known model of extreme value theory: the Peaks-over-Threshold, or POT. According to this approach, excesses over a high threshold follow a Generalized Pareto distribution, as stated by the Pickands, Balkema and de Haan's Theorem \cite{Balkema74}, \cite{Pickands}, and the number of excesses over time follows a homogeneous Poisson process \cite{Falk}. If the last statement were verified for large armed conflicts, it would mean that  no particular trend can be observed, i.e. that the propensity of humanity to generate big wars has neither decreased nor increased over time.

In order to avoid problems with the armed conflicts of antiquity and possible missing data, we here restrict our attention on all events who took place after 1500 AD, i.e. in the last 515 years. As we have shown in Section \ref{eanalysis}, missing data are unlikely to influence our estimates of the shape parameter $\xi$, but they surely have an impact on the number of observations in a given period. We do not want to state that we live in a more violent era, simply because we miss observations from the past.

If large events, those above the 25k threshold for raw data (or the 145k one for rescaled amounts), follow a homogeneous Poisson process, in the period 1500-2015AD, their inter-arrival times need to be exponentially distributed. Moreover, no time dependence should be identified among inter-arrival times, for example when plotting an autocorrelogram (ACF).

\begin{figure}[!htb]
\begin{subfigure}{.5\textwidth}
\includegraphics[width=\linewidth]{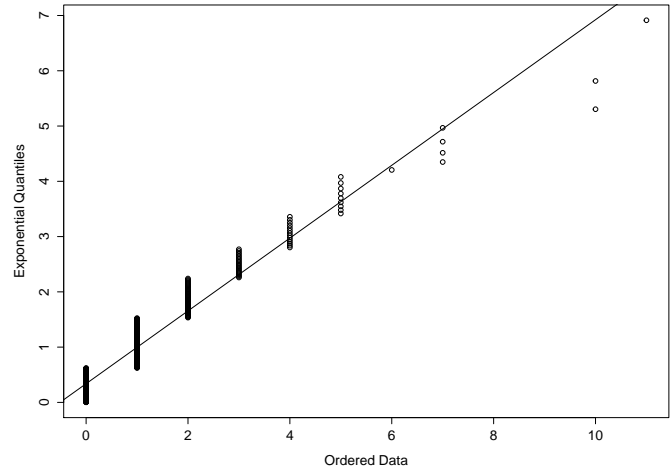}
\caption{Exponential QQ-plot of gaps (notice that gaps are expressed in years, so that they are discrete quantities).}
\label{QQgaps}
\end{subfigure}
\begin{subfigure}{.5\textwidth}
\includegraphics[width=\linewidth]{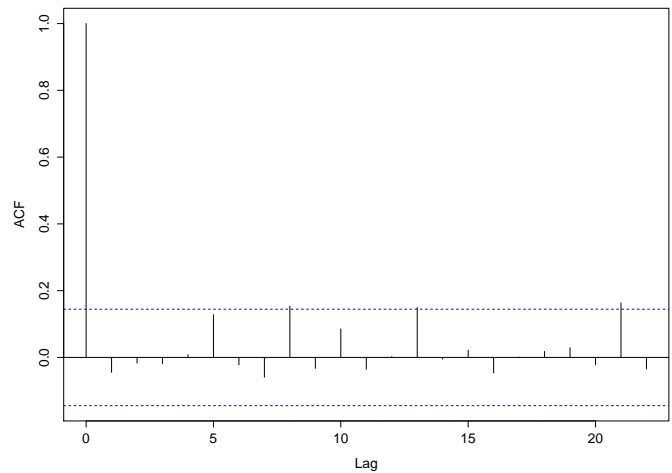}
\caption{ACF of gaps. Notice that the first lag has order 0.}
\label{ACF}
\end{subfigure}
\caption{Two plots to verify whether armed conflicts (raw data) over the 25k casualties threshold follow a homogeneous Poisson Process}
\label{Hypotheses}
\end{figure}

Figure \ref{Hypotheses} shows that both of these characteristics are satisfactorily observable in our data set.\footnote{When considering inter-arrival times, we deal with integers, as we only record the year of each event.} This is clearly visible in the QQ-plot of Subfigure \ref{QQgaps}, where most inter-arrival times tend to cluster on the diagonal.

Another way to test if large armed conflicts follow a homogeneous Poisson process is by considering the number of events in non-overlapping intervals. For a Poisson process, given a certain number of events in a time interval, the numbers of events in non-overlapping subintervals follow a Multinomial distribution. If the Poisson process is homogeneous, then the Multinomial distribution is characterized by the same probability of falling in any of the sub-intervals, that is an equiprobability. It is not difficult to verify this with our data, and to see that we cannot reject the null hypothesis of equiprobable Multinomial distribution, over the period 1500-2015, in which 504 events took place, choosing a confidence level of 10\%. 

Once again, the homogeneous Poisson behavior is verified for raw, rescaled and dual data. Regarding the estimates of $\xi$, if we restrict our attention on the period 1500-2015, we find $1.4093$ for raw data, and $1.4653$ for rescaled amounts.

To conclude this section, no particular trend in the number of armed conflicts can be traced. We are not saying that this is not possible over smaller time windows, but at least in the last 500 years humanity has shown to be as violent as usual. 

To conclude our paper, one may perhaps produce a convincing theory about better, more peaceful days ahead, but this cannot be stated on the basis of statistical analysis ---this is not what the data allows us to say. Not very good news, we have to admit.

\section*{Ackowledgements}
Captain Mark Weisenborn engaged in the thankless and gruesome task of initially compiling the data, checking across sources and linking each conflict to a narrative. We also benefited from generous help on social networks where we put data for scrutiny, as well as advice from historians such as Ben Kiernan.

We thank Raphael Douady, Yaneer Bar Yam, Joe Norman, Alex(Sandy) Pentland and his lab for discussions and comments.

Pasquale Cirillo acknowledges the support of his Marie Curie CIG ``Multivariate Shocks" (PCIG13-GA-2013-618794).

\end{document}